\documentclass[]{spie}
\usepackage[]{graphicx}
\usepackage{subfigure,amsmath, amssymb,amsfonts}
\usepackage[colorlinks=true, allcolors=blue]{hyperref}

\title{GPI 2.0: Upgrades to the IFS including new spectral modes}

\author[a]{Mary Anne Limbach}
\author[b]{Jeffrey Chilcote}
\author[c]{Quinn Konopacky}
\author[d]{Robert De Rosa}
\author[b]{Randall Hamper}
\author[e]{Bruce Macintosh}
\author[f]{Christian Marois}
\author[g]{Marshall Perrin}
\author[h]{Dmitry Savransky}
\author[f]{Jean-Pierre Veran}
\author[i]{Jason Wang}
\author[e]{Arlene Aleman}
\affil[a]{Department of Physics and Astronomy, Texas A\&M University, 4242 TAMU, College Station, TX 77843-4242 USA}
\affil[b]{Department of Physics, University of Notre Dame, 225 Nieuwland Science Hall, Notre Dame, IN, 46556, USA}
\affil[c]{Center for Astrophysics and Space Science, University of California San Diego, La Jolla, CA 92093, USA}
\affil[d]{European Southern Observatory, Alonso de Cordova 3107, Vitacura, Santiago, Chile}
\affil[e]{Kavli Institute for Particle Astrophysics and Cosmology, Stanford University, Stanford, CA 94305, USA}
\affil[f]{National Research Council of Canada Herzberg, 5071 West Saanich Road, Victoria, BC V9E 2E7, Canada}
\affil[g]{Space Telescope Science Institute, Baltimore, MD 21218, USA}
\affil[h]{Sibley School of Mechanical and Aerospace Engineering, Cornell University, Ithaca, NY 14853, USA}
\affil[i]{Department of Astronomy, California Institute of Technology, Pasadena, CA 91125, USA}

\begin{document}
\maketitle

\begin{abstract}
The Gemini Planet Imager (GPI) is a high-contrast imaging instrument designed to directly image and characterize exoplanets. GPI is currently undergoing several upgrades to improve performance. In this paper, we discuss the upgrades to the GPI IFS. This primarily focuses on the design and performance improvements of new prisms and filters. This includes an improved high-resolution prism which will provide more evenly dispersed spectra across y, J, H and K-bands. Additionally, we discuss the design and implementation of a new low-resolution mode and prism which allow for imaging of all four bands (y, J, H and K-bands) simultaneously at R$\approx$10. We explore the possibility of using a multiband filter which would block the light between the four spectral bands. We discuss possible performance improvements from the multiband filter, if implemented. Finally we explore the possibility of making small changes to the optical design to improve the IFS's performance near the edge of the field of view.
\end{abstract}

\section{Introduction}

The Gemini Planet Imager's (GPI's) science instrument, an integral field spectroscopy (IFS) has successfully discovered and taken spectra of many exoplanets and circumstellar disks (Macintosh et al. 2018\cite{Macintosh2018}) from the southern hemisphere over the past decade. GPI is now undergoing upgrades (Chilcote et al. 2018, 2020\cite{Chilcote2018, Chilcote2020}) to improve its performance and will be reinstalled on Gemini north in the near future. In this paper, we discuss upgrades to the GPI IFS including improvements to the high-resolution spectral mode (Section \ref{HRmode}) the addition of a new low-resolution spectral mode (Section \ref{LRmode}), and possible upgrades to the GPI IFS optics (Section \ref{ImprovedOpt}).

The GPI IFS is a near-infrared ($0.9-2.4\mu m$) lenslet-based instrument with a $2.7''\times2.7''$ field of view capable of detecting exoplanets within $<0.5 arcseconds$ of their host star. The IFS has a plate scale of 14.17mas/lenslet and makes use of a HAWAII-2RG detector. The original GPI IFS contained five bandpasses (Y, J, H,
K1 and K2). Analysis during the original GPI design showed that observations in H band alone at $R\sim40$ would provide sufficient accuracy in estimating temperatures and gravities of exoplanets, so that was the target resolution for H band. However, with obtainable materials, the prism does not provide this uniformly across the YJHK window (Figure 2). K band was split to allow the spectra to fit on the detector, but the new upgraded IFS will be capable of imaging all of K-band simultaneously rather than splitting it into K1 and K2 due to a new dispersion element. Further the upgraded spectral resolving power now varies less across the four bands from $R = \lambda/\Delta\lambda = \sim 46-77$. The IFS optics are in a dewar at $T = 70K$, $p = 1\times10^{-6}$ torr. The original GPI IFS design is furthered detailed in Chilcote et al. 2012\cite{Chilcote2012} and Larkin et al. 2014\cite{Larkin2014}.

\begin{figure}
\centering
\includegraphics[width=0.98\textwidth]{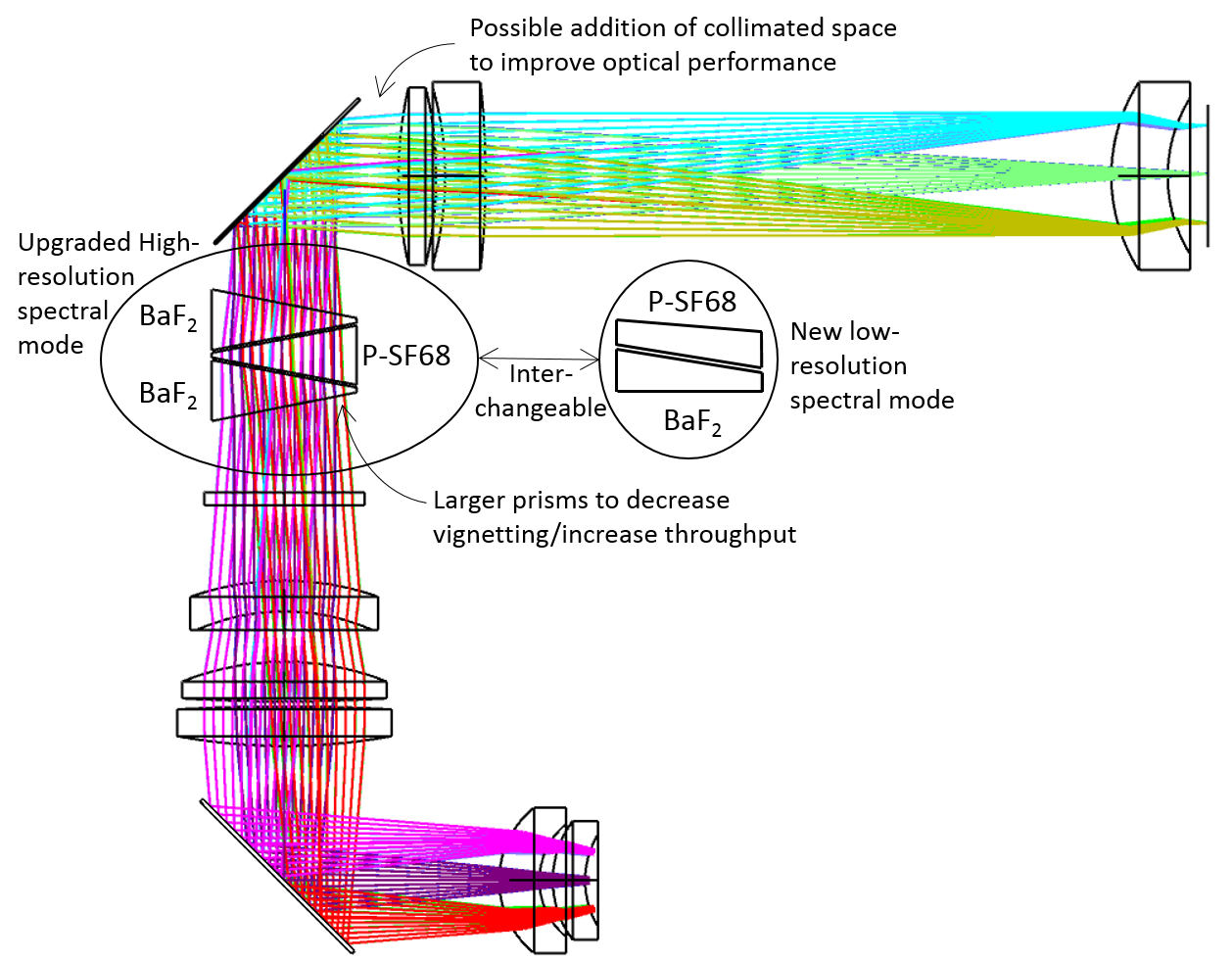}
\caption{Optical layout of the GPI IFS from the lenslet array (bottom focus) to detector (top right focus) showing where the upgrades and changes to the design will occur. These changes include (1) a new low-resolution spectral mode, (2) an upgraded high-resolution mode, (3) addition of collimated space and (4) larger prisms to decrease vignetting.}
\label{opticalLayout}
\end{figure}

The optical layout in Figure \ref{opticalLayout} shows the layout of the GPI IFS from post-lenslet focus (bottom) to detector focus (top right). The layout illustrates where the new design changes and additions will occur. Light is focused by the lenslet array (not shown) a the bottom of the opical layout. Five powered optics and a fold mirror are then used to collimated the F/3.52 beam. A filter and zero-deviation prism is placed in the collimated space to disperse each of the $\sim36,000$ PSF-lets created by the lenslet array. The prisms and filters are interchangeable to allow for the various GPI observing modes. A fold mirror followed by three optics then focus the light on to the H2RG detector (top right). Not shown here is the Wollaston prisms used for the GPI polarization mode. This optic will remain unchanged during the upgrade -- the addition of the low resolution spectral mode will be placed in an entirely new prism slot in the GPI IFS dewar.

\section{Upgraded High-Resolution Spectral Mode}\label{HRmode}
One of the goals of the GPI upgrade is to redesign the IFS prism to allow for simultaneous imaging of all of K-band and to produces spectra in each of the four bands that are roughly equal length. We conducted a design study to find the best combination of glass materials by examining the combined dispersion profiles of dozens of crown and flint glasses across the near infrared.

\begin{figure}
\centering
\includegraphics[width=0.85\textwidth]{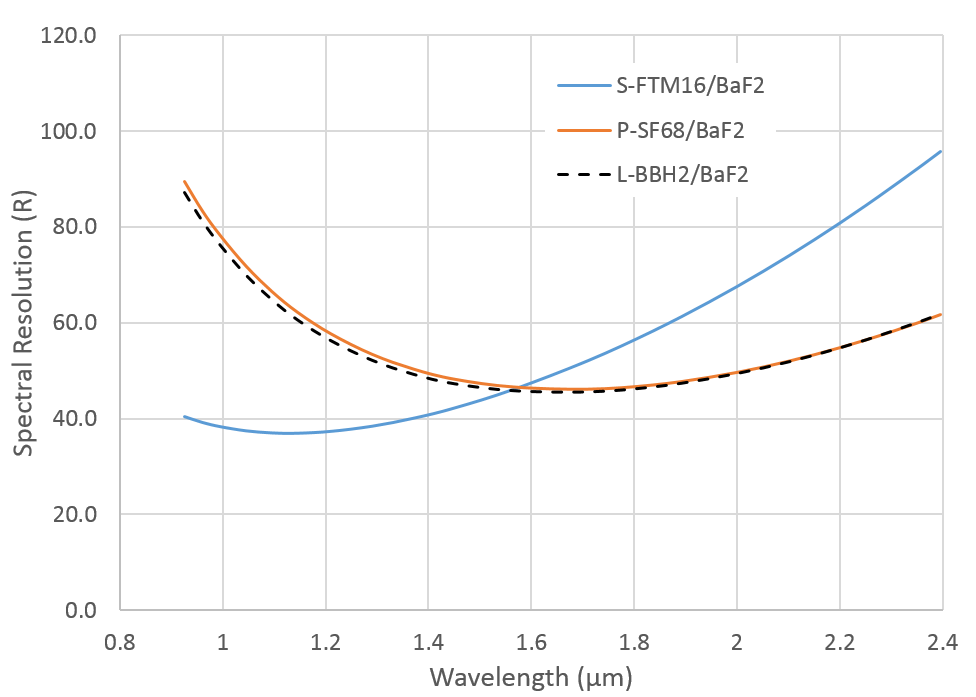}
\caption{Spectral dispersion (for high resolution prisms) of $P-SF68$/$BaF_2$, $L-BBH2$/$BaF_2$ and $S-FTM16$/$BaF_2$. Higher spectral resolution is desired at shorter wavelengths in order to achieve equal spectral length in y, J, H and K band. The $P-SF68$ (proposed for GPI upgrade) and $L-BBH2$ (currently used in Subaru/CHARIS) prisms provide near equal spectral lengths across all four bands. The $S-FTM16$/$BaF_2$ prism is currently used in GPI.}
\label{3prismsRes}
\end{figure}

The new high resolution prism is three elements: $P-SF68$ and two $BaF_2$ elements. This material combination was chosen to provide near-uniform spectral lengths of all four infrared bands as well as zero-deviation (see Table 2). We explored dozens of glass materials with similar index and abbe numbers and concluded that $P-SF68$ was optimal both due to its availability in relatively large blanks and it's resulting spectral dispersion profile. We compared the performance of $P-SF68$/$BaF_2$ to $L-BBH2$/$BaF_2$. $L-BBH2$ is no longer being manufactured, so it can not be used in GPI but is a good point of comparison because it provides uniform dispersion across all four bands and has been used in other high-contrast imaging IFSs (i.e. Subaru's CHARIS, Peters-Limbach et al. 2013\cite{Limbach2013}). A comparison of three zero deviation prisms ($P-SF68$/$BaF_2$, $L-BBH2$/$BaF_2$ and $S-FTM16$/$BaF_2$ (GPIs old high resolution prism), are shown in Figure \ref{3prismsRes}. Higher spectral resolution is desired at shorter wavelengths in order to achieve equal spectral length in all four bands. Note that the spectral resolutions shown here are for the high-resolution mode. We found that while $L-BBH2$ and $P-SF68$ provide nearly the same dispersion profile, larger angles (thicker prisms) are required to achieve the same dispersion with $P-SF68$/$BaF_2$ verses $L-BBH2$/$BaF_2$. However, P-SF68 is available in relatively large slabs, so we were able to design a 3-element $P-SF68$/$BaF_2$ prism was still sufficient to obtain our required dispersion. The front/back surfaces of the new prisms are $54\times54mm^2$ with an 88\% clear aperture. This is larger than the original high-resolution prism in GPI and should decrease vignetting and increase throughput from the prisms. 

\begin{center}
Table 1: GPI High-Resolution Spectral Mode Parameters
\begin{center}
\begin{tabular}{ c | c | c | c}
 Band & Cut-on/-off ($\mu m$) & Length (pixels) & $R = \lambda/\Delta\lambda$\\
 \hline 
y-band  & 0.95-1.07 & 18.3 & 76.9\\ 
J-band & 1.13-1.34 & 19.4 & 57.0  \\
H-band & 1.498-1.796 & 16.8 & 46.5 \\
K-band & 2.00-2.40 & 20.0 & 55.1 \\
\end{tabular}
\end{center}
\end{center}

The performance of the new GPI high-resolution prism are shown in Table 1. Most notably, the old GPI K-band spectral resolution was $R>70$ and is now $R\sim55$ which allows for all of K-band to now be imaged simultaneously. Further, the spectral resolution in y and J-bands both increased from $R\sim35$ to $R\sim57$ and $R\sim77$, respectively. These key changes in spectral resolution produce spectra of roughly the same length ($l = 18.4\pm1.6$) in each band on the detector. In addition to the filter change to one K-band, the y-band and J-band filters will also be upgraded in the new instrument. These filters will now be slightly narrower, but still include the entirety of what is typically defined as y and J-band. The previous filters were much broader than the typical y and J-band due to the low dispersion of the old prism at short wavelengths. The H-band filter will remain unchanged in the upgraded IFS. The exact as-built bandpass of the H-band filter and all of the new band cut-on/offs, spectral lengths and resolutions are listed in Table 1.

\section{New Low Resolution Spectral Mode}\label{LRmode}
\begin{figure}
\centering
\includegraphics[width=0.8\textwidth]{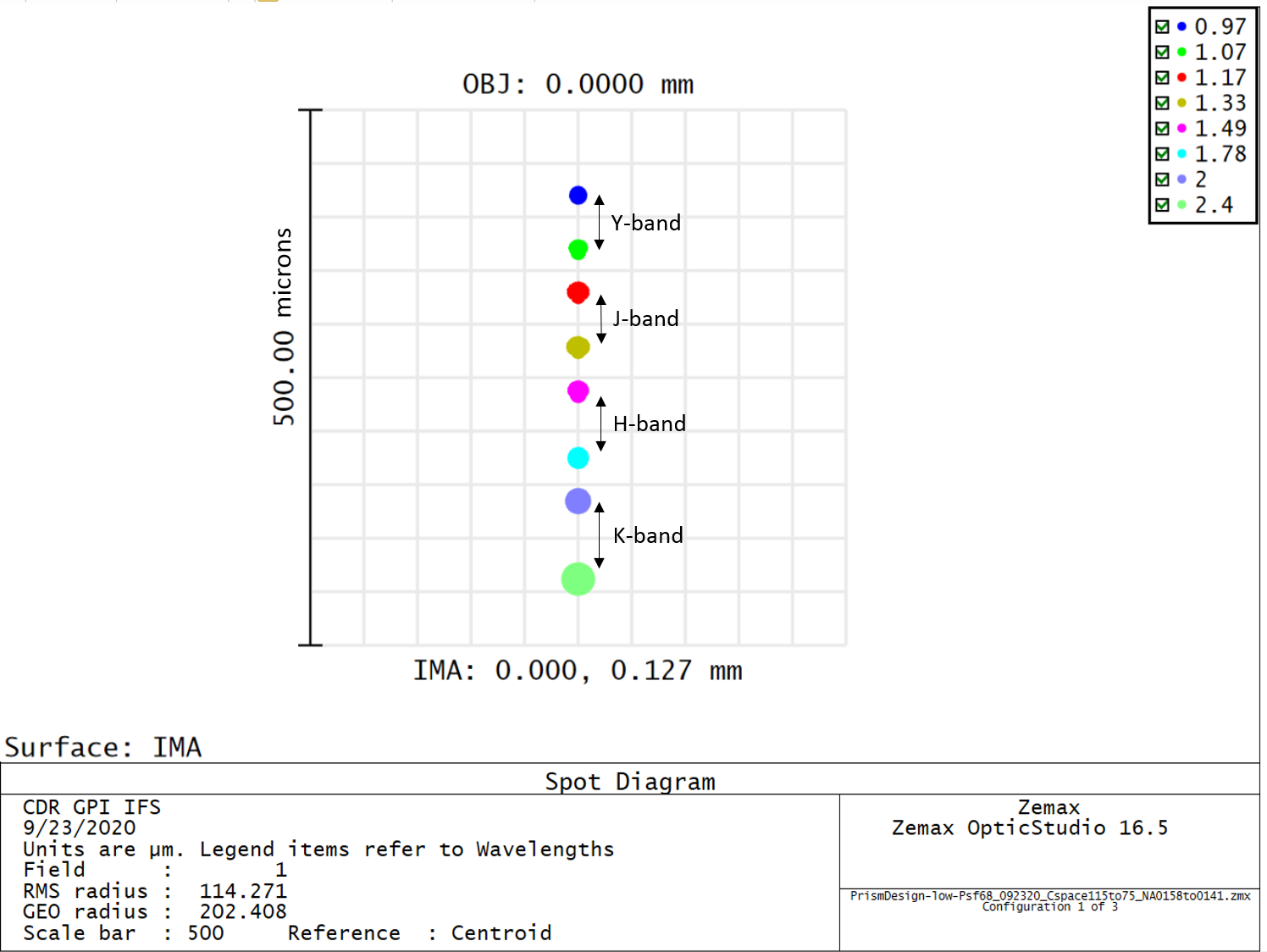}
\caption{On-axis spot diagrams in low-resolution spectral mode at the detector. This figure helps visualize the size and spectral resolution of an individual spectrum in the low resolution spectral mode. For reference, one pixel is 18 microns.}
\label{LRspots}
\end{figure}

As part of the upgrades to GPI, we are implementing a new low-resolution mode that will allow for simultaneous imaging across y, J, H and K band ($\lambda = 0.97-2.40 \mu m$). This is similar to the low resolution mode on Subaru's CHARIS (Peters-Limbach et al. 2013\cite{Limbach2013}), which has proven useful for both detection and characterization of exoplanets (Currie et al. 2018\cite{Currie2018}, 2019\cite{Currie2019}, Rich et al. 2019\cite{Rich2019}, Wang et al. 2020\cite{Wang2020}). Although the original GPI design specified $R\sim40$ to estimate temperatures accurately, the experience with CHARIS and simulations by the GPIES collaboration indicate that (for a fixed exposure time) the broader wavelength coverage of this low-resolution mode offsets the lower spectral resolution and can still provide accurate temperature estimates (particularly useful for initially distinguishing planets from background objects.) Also, the GPI 2.0 science case emphasizes fainter target stars and planets, and the lower dispersion will reduce the impact of readout noise on survey observations. The new mode will provide a spectral resolution of $R = \lambda/\Delta\lambda \approx 10$ across all bands (roughly 2-4 pixels per band). The total spectral length is 20 pixels with a gap between each spectra of 2.5 pixels on the detector. Table 2 lists the number of pixels across each band and the corresponding spectral resolution. Note that there is a 2-3 pixel gap between each band. Figure \ref{LRspots} shows the on-axis spot diagram of one spectra on the detector.

\begin{center}
Table 2: GPI Low-Resolution Spectral Mode Parameters
\begin{center}
\begin{tabular}{ c | c | c | c}
Cut-on/-off ($\mu m$) & Band & Length (pixels) & $R = \lambda/\Delta\lambda$\\
 \hline 
0.97 - 1.07 & y-band  & 2.9 & 15.0\\ 
 & GAP & 2.3 & \\
1.17 - 1.33 & J-band & 2.8 & 11.1 \\
 & GAP & 2.2 & \\
1.49 - 1.78 & H-band & 3.3 & 9.3 \\
 & GAP & 2.3 & \\
2.0 - 2.4 & K-band & 4.1 & 11.2 \\
 & {\it Total length:} & 20.0 \\
\end{tabular}
\end{center}
\end{center}

The new low resolution prism is two elements: $P-SF68$ and $BaF_2$. These prism materials were chosen by the same design study described in the previous section. For the relatively small amount of dispersion required for the GPI low-resolution mode, a low profile 2-element $P-SF68$/$BaF_2$ prism was sufficient to obtain our required dispersion.  The normal surfaces of the new prisms are $50.85\times50.85mm^2$ with an 88\% clear aperture. This is larger than the original high-resolution prism in GPI to decrease vignetting, however, it is slightly smaller than the new high-resolution prism because the prisms low-profile allows it to be placed closer to the pupil plane.

The new low-resolution spectral mode will improve GPI's ability to search for new exoplanets. Because light can be collected from all four bands simultaneously in this mode, less total observation time is required to detect an exoplanet. This implies that a larger number of stars can be surveyed in less time increasing the likelihood of new exoplanet discoveries.

The nominal filter to be used in the low resolution spectral mode will transmit all light from $0.97-2.40 \mu m$, including the light in the gaps between the bands. Because the Earth's atmosphere does have some transmission in these bands it is possible that the light detected in these gaps between the bands can be used to detect water in the atmospheres of exoplanets. However, such observations will be challenging due to variations in Earth's water vapor column, sky background and low transmission through Earth's atmosphere in these gaps.

\begin{figure}
\centering
\includegraphics[width=1\textwidth]{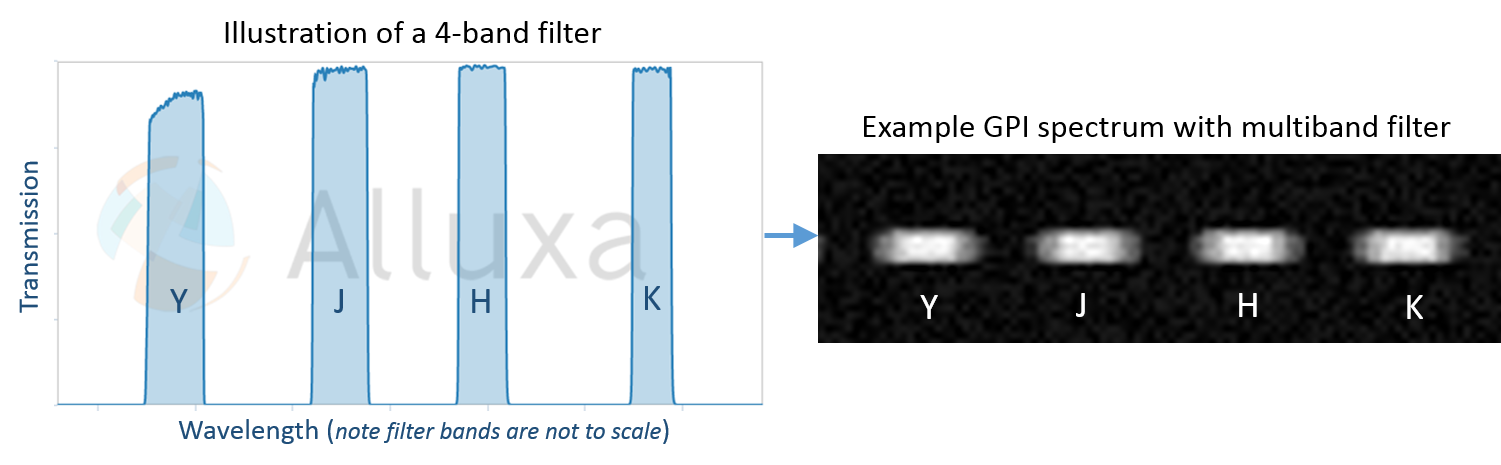}
\caption{Illustration of a 4-band filter (left) that could be implemented with the low-resolution prism. The resulting spectrum on the GPI detector (right) is four well separated images of the science target in each spectral band. The well separated spots can provide an independent means of wavelength calibration as well as other performance improvements. Note that the filter shown on the left is manufacturable but this is just an plot of what a different existing Alluxa Inc. quad-band filter looks like with GPI's four spectral bands overlaid to demonstrate the idea - this new filter is not yet designed or built.}
\label{4bandFilt}
\end{figure}

In addition to this nominal low-res operating mode, GPI is considering implementing a multiband y+J+H+K filter. The filter could improve GPI's performance (for the reasons discussed below) and provide isolated images of each of the bands. This filter would only transmit the light in the four bands and block the light in the gaps. A illustration of this filter and corresponding spectral image at the detector are shown in Figure \ref{4bandFilt}.  Multiband filters, such as this 4-band filters, require relatively new thin-film coating technology, but are reliably manufactured by companies such as Alluxa, Inc. with very high ($>95\%$) average in-band transmission. If we choose to implement this 4-band filter, we expect several improvements to GPI's performance. First, some of the sky background noise from the gaps leak into adjacent bands. Such a filter would block this unwanted light decreasing noise. Also, the large number of edges on this filter could potentially be used to determine the wavelength solution of the GPI spectra. Currently GPI must rely on an external source to measure the wavelength solution and this procedure introduces some amount of systematic noise (Wolff et al. 2014\cite{Wolff2014}). If the 4-band filter is implemented, every spectrum in the GPI image (exoplanet, star and calibration spots) will have eight distinct filter edges which can be used to solve the wavelength solution without an external source. 

Finally, it is possible that implementing a multi-band filter could provide a new method for measuring colors which could improve relative color photometry measurements in GPI. Currently GPI makes color photometry measurements by referencing the exoplanet spectrum to satellite spots. However, this method is subject to many systematic errors and as such the source of noise limiting GPI's photometric precision is currently poorly understood. The multiband filter + prism create well-separated spots/PSFs (just like the Wollaston prism in polarization mode). Therefore it is possible that rather than referencing the exoplanet spectrum to the satellite spots to make a color measurement, one could instead reference one band of the exoplanet to another band of the exoplanet to make a relative color measurements. This is akin to the common-path multiband imaging technique described in Limbach et al. in prep.\cite{Limbach2021} This method is used to obtain relative measurements in GPI's polarization mode. GPI's polarization mode is capable of $\sim1\%$ relative precision, $5-10\times$ better than GPI's current photometric precision. This indicates that implementing this method in spectral mode may improve the photometric precision. However, the extraction process for exoplanets introduces systematic noise (which is poorly understood). It is possible that this extraction method will remain the dominant source of noise in GPI's photometric precision even if this new mode is implemented. Despite this unknown, there are still other clear advantages discussed previously that suggest the addition of a 4-band filter in GPI will provide some improvement.

\section{Improved Optical Performance}\label{ImprovedOpt}
\begin{figure}[]
\centering
\includegraphics[width=1\textwidth]{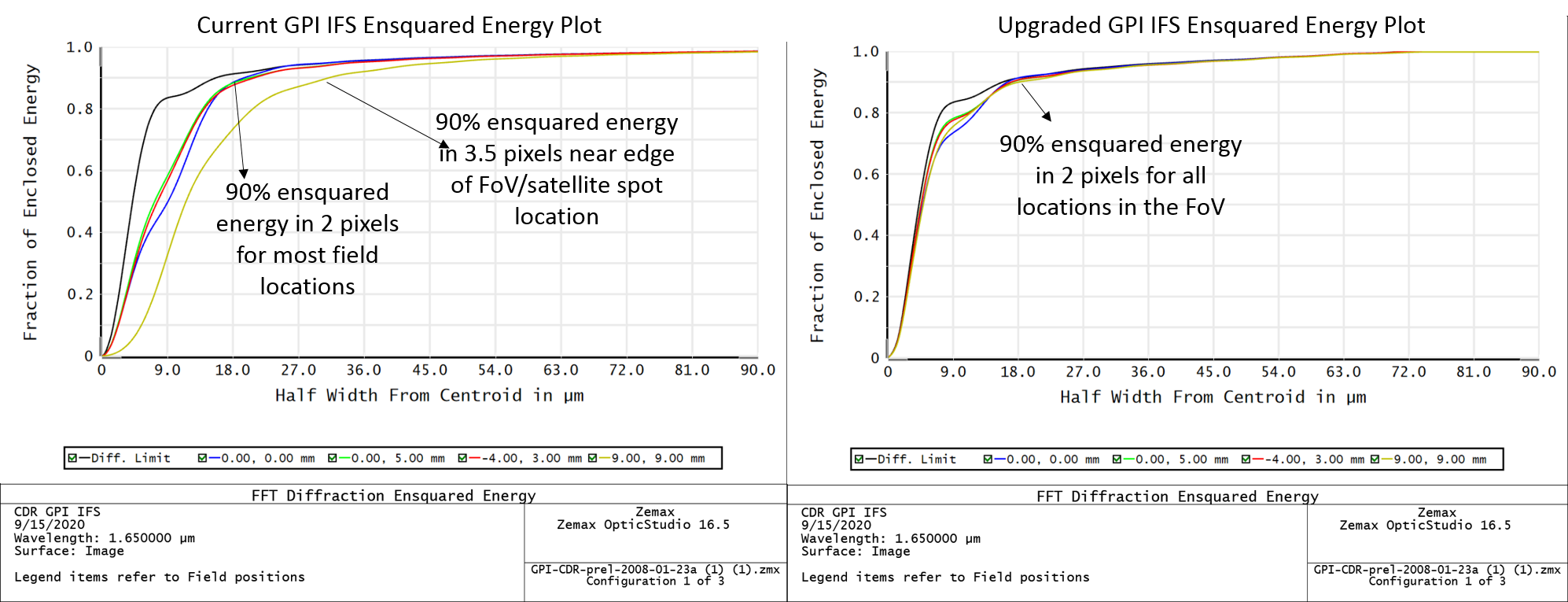}
\caption{Ensquared energy at various positions on the detector for the current GPI instrument (left) and with proposed upgrades (increased collimated space, right). For reference, one pixel is 18 microns.}
\label{EEplot}
\end{figure}

The edge of GPI's field of view suffers from some astigmatism and coma. We found that a slight increase (+20mm) in the amount of collimated space will drastically improve the performance at the edge of GPI's field of view (see Figure \ref{EEplot}). The plot shows that currently near the edge of the field 3.5 pixels ensquare 90\% of the light whereas the image is much sharper (90\% of the ensquared energy in 2 pixels). By increasing the collimated space, the IFS will become diffraction limited across the field of view with 90\% ensquared energy in 2 pixels for all field positions. We are currently exploring the possibility of increasing the collimated space in IFS. This improvement in performance is particularly important because the satellite spots, which are used for calibration lie near the edge of the field and are subject to these optical aberrations. If this improvement is implements, calibrations and extractions are like to be more precise which will in turn improve GPI's spectral and photometric precision.

\section{Conclusion}
In this paper we discussed the design and performance of the new GPI IFS low-resolution spectral mode which will allow for simultaneous imaging across y, J, H and K-band. Further, we discussed the possibility of including a 4-band filter in the low-resolution mode and the performance and calibration benefits associated with this. We also detailed the design and performance of the upgraded version of the high-resolution mode which will provide more uniform dispersion across all four bands and will allow GPI to combine K1 and K2 into one K-band spectral observing mode. Finally, we considered the possibility of increasing collimated space in the GPI IFS to improve optical performance near the edge of GPI's field of view. All together these upgrades will prove GPI's capabilities and performance on-sky making more likely to detect and image exoplanets and easier to characterize their atmospheres.

\section*{Acknowledgments}
The GPI 2.0 instrument upgrades are funded by NSF MRI Award Number 1920180. The GPI project has been supported by Gemini Observatory, which is operated by AURA, Inc., under a cooperative agreement with the NSF on behalf of the Gemini partnership: the NSF (USA), the National Research Council (Canada), CONICYT (Chile), the Australian Research Council (Australia), MCTI (Brazil) and MINCYT (Argentina).

{}

\end{document}